 \documentclass[twocolumn,twoside,showkeys,showpacs,preprintnumbers,superscriptaddress,amsmath,amssymb,prc]{revtex4-2}

\usepackage{graphicx}
\usepackage{dcolumn}
\usepackage{bm}
\usepackage[version=4]{mhchem}
\usepackage{subcaption}
\usepackage[colorlinks=true,linkcolor=black,citecolor=blue,urlcolor=blue,]{hyperref}
\usepackage{appendix}
\usepackage{balance}
\usepackage[justification=raggedright,singlelinecheck=false]{caption}

\DeclareUnicodeCharacter{0394}{$\Delta$}
\DeclareUnicodeCharacter{03B3}{$\gamma$}

\def\thu{Department of Physics, Tsinghua University, Beijing 100084, China}
\def\nju{School of Physics, Nanjing University, Nanjing 210093, China}

\begin{document}


\title{Reconstruction of Bremsstrahlung $\gamma$-rays Spectrum in Heavy Ion Reactions with Richardson-Lucy Algorithm}

\author{Junhuai Xu}
\email{xjh22@mails.tsinghua.edu.cn}
\affiliation\thu
\author{Yuhao Qin}%
\affiliation\thu

\author{Zhi Qin}
\email{qinz18@mails.tsinghua.edu.cn}
\affiliation\thu

\author{Dawei Si}
\affiliation\thu


\author{Boyuan Zhang}
\affiliation\thu

\author{Yijie Wang}
\affiliation\thu

\author{Qinglin Niu}
\affiliation\nju

\author{Chang Xu}
\affiliation\nju


\author{Zhigang Xiao}
\email{xiaozg@tsinghua.edu.cn}
\affiliation\thu



\date{\today}

\begin{abstract}
The high momentum tail (HMT) in the momentum distribution of nucleons above the Fermi surface has been regarded as an evidence of short-range correlations (SRCs) in atomic nuclei. It has been showcased recently that the $np$ Bremsstrahlung radiation in heavy ion reactions can be used  to extract HMT information. The Richardson-Lucy (RL) algorithm is introduced to the reconstruction of the original Bremsstrahlung $\gamma$-ray energy spectrum from experimental measurements. By solving the inverse problem of the detector response to the $\gamma$-rays, the original energy spectrum of the Bremsstrahlung $\gamma$ in 25 MeV/u $^{86}$Kr + $^{124}$Sn has been reconstructed and compared to the isospin- and momentum-dependent Boltzmann-Uehling-Uhlenbeck (IBUU) simulations. The analysis based on hypothesis test suggests the existence of the HMT of nucleons  in nuclei, in accordance with the previous conclusions. With its effectiveness being demonstrated, it is feasible to apply the RL algorithm in future experiments of measuring the Bremsstrahlung $\gamma$-rays in heavy ion reactions.

\end{abstract}

\keywords{High Momentum Tail, Bremsstrahlung $\gamma$-ray, Richardson-Lucy Algorithm, Hypothesis Test}

\pacs{ }

\maketitle


\section{Introduction}

The strong interaction between  nucleon pairs at short distance with high relative momentum  in nucleus is called short-range correlation (SRC). Because of the SRC effect, the high momentum tail (HMT) in the momentum distribution of the nucleons in nuclei above the Fermi surface exists \cite{RevModPhys.89.045002} and can be measured experimentally. So far much of our  knowledge on the SRC, which gives suggestively rise to the  EMC effect \cite{EuropeanMuon:1983wih}, is obtained from high-energy electron-nucleus scatterings \cite{Hen:2014nza,CLAS:2018yvt,CLAS:2019vsb,Li:2022fhh} and from proton induced reactions \cite{Tang:2002ww,Piasetzky:2006ai}. Very recently, it was pointed out that the existence of the HMT nucleons may harden the energy spectrum of the energetic  photons emitted in $np$ Bremsstrahlung process in heavy ion reactions at intermediate energies \cite{XUE2016486}.  

Because of the merit that the energetic photons experience rare final interactions with the surrounding nuclear medium once they are produced, experimental measurement of the Bremsstrahlung photons follows up. In our previously work \cite{QIN2024138514},  the energy spectrum up to 80 MeV of the Bremsstrahlung $\gamma$-rays has been measured and analyzed using the isospin- and momentum-dependent Boltzmann-Uehling-Uhlenbeck (IBUU) simulations.  Despite of the insufficient statistics and the high energy border of $E_\gamma<80$ MeV, the experimental results suggest the  existence  of HMT at a confidential level of about 90\% and showcase the feasibility to study SRC by the Bremsstrahlung photons. To improve the confidential level, further experimental efforts are required. 

However, there is a hindrance. When comparing the experimental results to the theoretical calculation incorporating the HMT, the detector filtering effect must be taken into account. This is not trivial because the high energy $\gamma$-rays undergo complicated transport process in the detection system. Usually, it is convenient to solve the forward problem, i.e., to convert the theoretic prediction to the expected spectrum by the simulation of the detector response.  From a practical point of view, it then hinders  further comparative studies unless the accompanying response matrix of the detection system is always provided. Thus, it is favorable to solve the inverse problem, i.e., to reconstruct the original spectrum from the measured one.        

The Richardson–Lucy (RL) algorithm \cite{Richardson:72,1974AJ.....79..745L} based on Bayesian theorem has been widely applied in optical imaging deblurring problems\cite{10.5120/20396-2697}, mainly to repair the blurred images with known point distributions. In recent years, it has seen  increasing applications in some nuclear physics problems. 
For instance, the velocity distribution of emitted particles measured experimentally can be used to infer the unknown velocity distribution of the center of mass system, thereby to determine the original momentum distribution of HIC products \cite{PhysRevC.105.034608}. The RL algorithm has also been applied in the analysis of particle correlation functions, from which one can  infer the distribution of the source functions \cite{NZABAHIMANA2023138247}. In the field of energy spectrum measurement and analysis, the RL algorithm has been proven practical in obtaining the shell structure of unbound particle systems from the measured decay energy spectrum \cite{PhysRevC.107.064315}. In the data analysis  of HIC experiments, statistical fluctuations can affect the stability of results, and RL algorithm is still an effective tool for optimizing the stability of solutions \cite{VARGAS201316}. In addition, during the iterative process on which the RL algorithm is running,  the output of each iteration maintains a positive value. For the problem of solving $\gamma$ spectra, the RL algorithm  avoids directly inverting the detector response matrix and efficiently solves the original $\gamma$ spectra.

Therefore, we are motivated to introduce the RL algorithm in the reconstruction of the original $\gamma$ spectrum, which can be compared directly to the model prediction. The  paper is arranged as following. In section \ref{secRL}, we describe the RL algorithm for solving the inverse problem in $\gamma$ measurement. In section \ref{secTest} we test and verify the accuracy of the RL algorithm. In section \ref{secanalysis}, the RL method is applied to reconstruct the original $\gamma$ spectrum  from the measured one in 25 MeV/u $^{86}$Kr + $^{124}$Sn  \cite{QIN2024138514}, and the comparison between the reconstructed  original $\gamma$ energy spectrum and the simulation of IBUU model is discussed.  Section V is the summary.

\section{Richardson-Lucy Algorithm}\label{secRL}

\subsection{Model Description}

In optical deblurring problems, a certain real property $\mu$ of a photon is measured as $\nu$ in experiments. The  distribution function $\mathcal {F}(\mu)$ of the real property $\mu$ and the distribution function $f(\nu)$ of the measured property $\nu$ is connected by the following equation,
\begin{equation}
    f(\nu)=\int d\mu P(\nu|\mu)\mathcal{F}(\mu).
\end{equation}
Here, $P(\nu|\mu)$ is the conditional probability of measuring photons with real properties $\mu$ as $\nu$.

Defining $Q(\mu|\nu)$ to represent the probability density of the real value being $\mu$ if $\nu$ is measured, which is the complementary probability density of $P$, one can connect $P(\nu|\mu)$ and $Q(\mu|\nu)$ by the Bayesian law.  Assuming that the measured value is within the range of $d\nu$, the probability within the range of $d\mu$ satisfies the following equation, 
\begin{equation}
    Q(\mu|\nu)f(\nu)d\nu d\mu=P(\nu|\mu)\mathcal{F}(\mu)d\mu d\nu.
\end{equation}

Equivalently,  $Q(\mu|\nu)$ can be expressed as,
\begin{equation}
\begin{aligned}
    Q(\mu|\nu)&=\frac{P(\nu|\mu)\mathcal{F}(\mu)}{f(\nu)}=\frac{P(\nu|\mu)\mathcal{F}(\mu)}{\int d\mu' P(\nu|\mu')\mathcal{F}(\mu')},\\
    \end{aligned}
\end{equation}
and $\mathcal {F}(\mu)$ can be expressed as
\begin{equation}
\mathcal{F}(\mu)=\frac{\int d\nu Q(\mu|\nu)W(\nu)f(\nu)}{\int d\nu' W(\nu')P(\nu'|\mu)},
\end{equation}
where $W(\nu)$ is the weight of the relative importance of specific data point at $\nu$ to infer $\mathcal{F}$. The weight of each measurement in the calculation can be manually specified to reduce the impact of unreliable data on the calculation. Meanwhile, by varying the $W$ one can  test the stability of the final solution obtained through calculation.

Use the RL method to iteratively solve the above two equations, one writes
\begin{equation}
    \begin{aligned}
        Q^{(r)}(\mu|\nu)&=\frac{P(\nu|\mu)\mathcal{F}^{(r)}(\mu)}{\int d\mu' P(\nu|\mu')\mathcal{F}^{(r)}(\mu')},\\
    \end{aligned}
\end{equation}
and 
\begin{equation}
    \begin{aligned}
        \mathcal{F}^{(r+1)}(\mu)&=\frac{\int d\nu Q^{(r)}(\mu|\nu)W(\nu)f(\nu)}{\int d\nu' W(\nu')P(\nu'|\mu)}.
    \end{aligned}
\end{equation}

Here the superscript  $r$ is the iteration index. It is seen that starting with a given initial distribution $ \mathcal{F}^{(0)}(\mu)$, one can obtain the final distribution after certain iteration times achieving convergence. The implicit dependency of $P$ on $\mathcal{F}$ can be handled in the iteration of the equation, and the non-negative distribution remains unchanged during the iteration process. By combining the above two equations \cite{PhysRevC.105.034608}, we can obtain

\begin{equation}
\begin{aligned}
    \mathcal{F}^{(r+1)}(\mu)&=\mathcal{F}^{(r)}(\mu)\frac{\int d\nu\frac{f(\nu)}{f^{(r)}(\nu)}W(\nu)P(\nu|\mu)}{\int d\nu' W(\nu')P(\nu'|\mu)}\\
    &=A^{(r)}(\mu)\cdot \mathcal{F}^{(r)}(\mu).
\end{aligned}
\end{equation}

Here, $A^{(r)}(\mu)$ is called the amplification factor, and $f^{(r)}(\nu)$ is the distribution function of $\nu$ obtained after the $r^{\rm th}$ iteration, that is
\begin{equation}
    f^{(r)}(\nu)=\int d\mu P(\nu|\mu)\mathcal{F}^{(r)}(\mu).
\end{equation}

\subsection{Application to the $\gamma$ Energy Spectrum Reconstruction}

In the experiment of HIC, $\gamma$ with a certain energy $\mu$ is measured to have energy $\nu$ through the response of the detector. Similar to optical deblurring problems, the RL algorithm can also be used to iteratively solve the original energy spectrum of $\gamma$.

The relationship between the distribution function $\mathcal{E}(\mu)$ of the discret real energy spectrum and the distribution function $e(\nu)$ of the measured energy spectrum is written in matrix form,
\begin{equation}\label{RelationshipExpresstion}
    e_i = \sum_jD_{ij}\mathcal{E}_j.
\end{equation}
Here, $D_{ij}$ is the detector response matrix, which can also be understood as the conditional probability that the $\gamma$ measurement with real energy $\mu$ is $\nu$. The values of  matrix elements $D_{ij}$ can be obtained generally by Geant4 simulation if the experimental setup and data analysis scheme are provided.

According to the RL formulation, one can reconstruct the original energy spectrum using the following iteration,
\begin{equation}
    \mathcal{E}_i^{(r+1)}=A_i^{(r)}\cdot \mathcal{E}_i^{(r)}.
\end{equation}
where $r$ in the above equation represents the iteration index, and the amplification factor $A_i^{(r)}$ after each iteration can be written as
\begin{equation}
    A_i^{(r)}=\sum_j\frac{e_j}{e_j^{(r)}}T_{ji},
\end{equation}
where $e^{(r)}$ is the value of the predicted measurement spectrum obtained in the $r^{\rm th}$ iteration, i.e.
\begin{equation}
    e_j^{(r)}=\sum_i D_{ji}\mathcal{E}_i^{(r)},
\end{equation}
$T_{ji}$ is the weighted energy spectrum transformation matrix, which has been automatically normalized as,
\begin{equation}
    T_{ji}=\frac{W_jD_{ji}}{\sum_{j'}W_{j'}D_{j'i}}.
\end{equation}
In HIC experiments, the manifestation of measuring energy spectrum is the counts of detector signals in different energy bins. The count distribution in each bin  satisfies the Poisson distribution. Accordingly, the weight $W$ of each measurement energy point used in iteration can be set to $\sqrt{e}$.

\section{Algorithm Test}\label{secTest}

In order to verify the effectiveness of the RL algorithm in solving the $\gamma$ original energy spectrum, the IBUU model was applied to calculate the  energy spectrum of the bremsstrahlung  $\gamma$ emission from HICs. The spectrum is then filtered by the detector response obtained by Geant4\cite{Geant4} simulations. Once the  energy spectrum from the detector output is generated, one can reconstruct inversely the input energy spectrum using the RL algorithm and compare to the original one.  The consistency is a proof of the effectiveness of the RL algorithm.

\begin{figure}[htbp]
\includegraphics[width=0.4\textwidth]{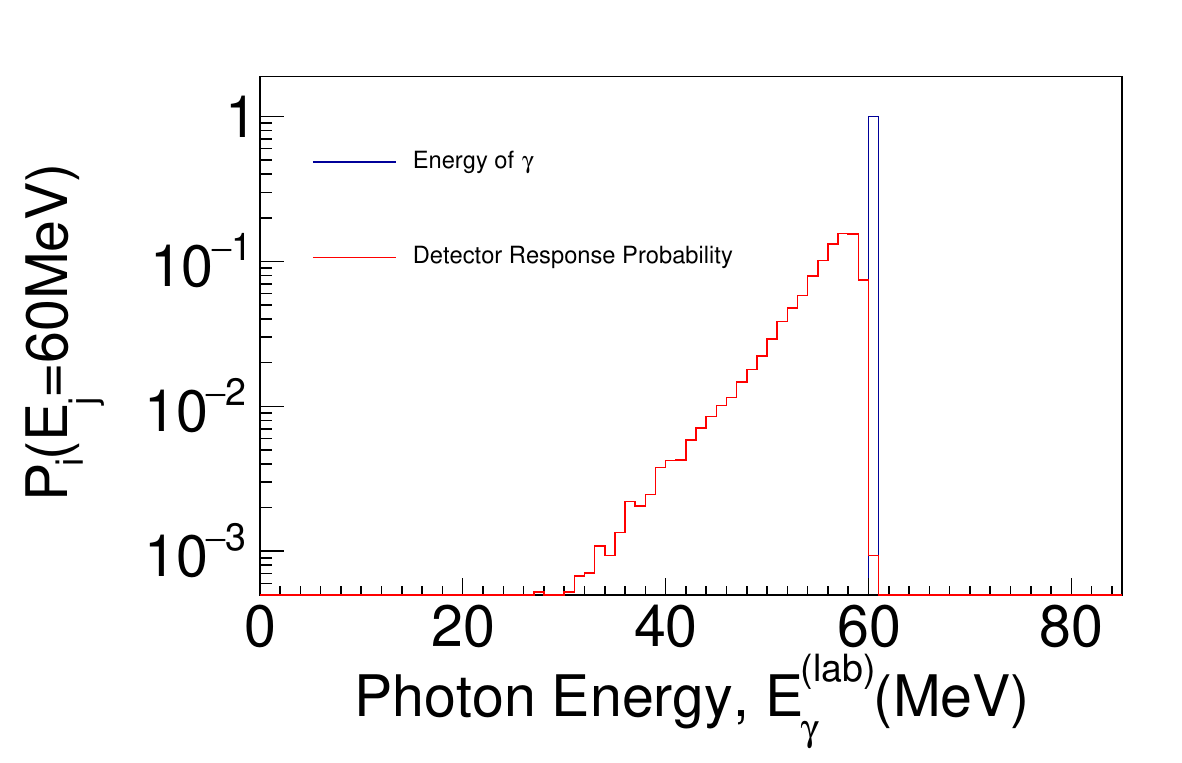}
\caption{\label{Energyresponse} (Color Online)  The response of $E^{(\rm lab)}_\gamma=60$ MeV single energy photons in the detector array. The energy on the abscissa is in laboratory reference frame.}
\end{figure}

For the test we take the 25  MeV/u $\rm {^{86}Kr + ^{124}Sn}$  reactions performed on the compact spectrometer for heavy ion experiment (CSHINE)  \cite{QIN2023168330,QIN2024138514}. A $4\times4$ CsI(Tl) hodoscope was mounted on CSHINE  to measure the high energy $\gamma$ rays \cite{QIN2023168330,GUAN2021165592}. The central $2\times 2$ units in the hodoscope were used to record the event center where the highest energy deposit was found. The total $\gamma$ energy was then collected from all neighbouring units fired with a given time window. For the details of the experiment, one can refer to  \cite{QIN2023168330}.
With this detector setup, one can simulate the detector response with Geant4 packages. As an example, Fig.~\ref{Energyresponse} shows the response of the detector caused by $E_\gamma=60$ MeV single energy photons. The red histogram presents the probability $P_i(E_j)$ of a single energy $E_j$ photon (here $E_j=60$ MeV indicated by the blue line) producing a response $E_i$ in the hodoscope, equivalently to say, $D_{ij}=P_i(E_j)$.

The BUU transport model \cite{Li:1996ix,Li:1997rc,Li:2003ts} effectively describes the dynamics of nucleus-nucleus collisions, with its main equation given by
\begin{equation}
    \frac{\partial f}{\partial t}+\mathbf{v}\cdot \nabla_r f-\nabla_r U\cdot \nabla_p f=I_{\rm coll},
\end{equation}
where $f(\mathbf{r},\mathbf{p},t)$ is the probability of finding a particle at time $t$ with momentum $\mathbf{p}$ at position $\mathbf{r}$, $U$ represents the mean-field potential, and the evolution  of $f(\mathbf{r},\mathbf{p},t)$ by elastic and inelastic two-body collisions is governed by the collision term  $I_{\rm coll}$. The original $\gamma$ spectrum is calculated by the IBUU transport model, in which the  SRC and HMT effects have been incorporated  \cite{LI201829,XUE2016486}. 
The simulations utilize the initial single-nucleon momentum distribution $n(k)$ that includes HMT. The behavior of HMT is remarkably consistent from the deuteron to infinite nuclear matter. The $n(k)$ for the HMT induced by SRC \cite{Hen:2014yfa} is given by
\begin{equation}
n(k)=\left\{
\begin{array}{rcl}
A, & & {k\le k_{\rm F}}\\
C/k^4, & & {k_{\rm F}<k\le \lambda k_{\rm F}}\\
0, & & {k>\lambda k_{\rm F}}
\end{array} \right.
\end{equation}
where $k$ represents the single nucleon momentum, $k_{\rm F}$ is Fermi momentum and $\lambda$ is the high-momentum cutoff. Due to the low probability of producing Bremsstrahlung photons in the reaction, the impact of bremsstrahlung on nucleon kinematics is also negligible, and hence, perturbation methods can be used to calculate the probability of photon production in each $np$ collision. Based on the single boson exchange (OBE) method, a well fitted expression for the probability of elementary double differential photon production can be applied in IBUU simulations \cite{Gan1994298},
\begin{equation}
    \frac{d^2P}{d\Omega dE_{\gamma}} = 1.67\times 10^{-7}\times\frac{[1-(E_{\gamma}/E_{\rm max})^2]^{\alpha}}{E_{\gamma}/E_{\rm max}}.
\end{equation}
Here, $E_{\gamma}$ represents the energy of produced photons, and $E_{\rm max}$ represents the total available energy in the center of mass frame. The coefficient $\alpha$ is $\alpha = 0.7319 - 0.5898\beta_i$, where $\beta_i$ represents the nucleon velocity. The total photon production probability per event is the sum of the probabilities of all $np$ collisions producing photons throughout the entire process.  From the total probability, one can obtain the original $\gamma$ spectrum with  given collision times $N_{\rm evt}$.

In our test, the stiffness  parameter $x$ of the symmetry energy is taken as  $x=-1$.  The triangular distribution of $b\le5$ fm is adopted to describe the collision centrality. Three ratios of the high momentum tail ($R_{\rm HMT}$) are adopted in the calculation,  including $R_{\rm HMT}=0\%$ as a free Fermi gas (FFG), $R_{\rm HMT}=15\%$, and $30\%$, respectively \cite{LI201829,XUE2016486}.

\begin{figure}[htbp]
\includegraphics[width=0.4\textwidth]{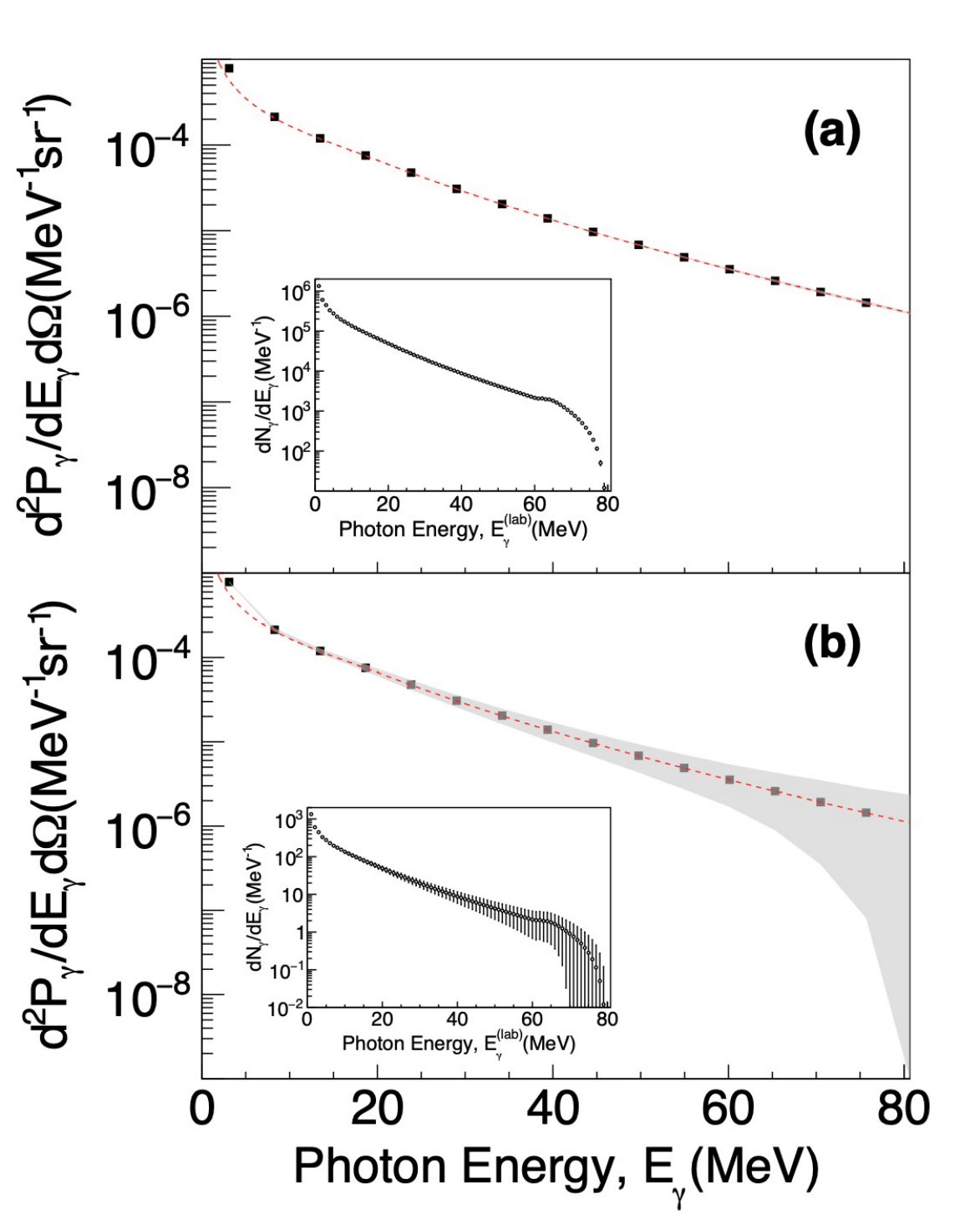}
\caption{\label{AlgorithmTestenough} (Color Online)  Comparison of the input and output spectra (in CM reference frame) of RL algorithm for a high-statistics (a) and low-statistics (b) test. The dashed curves are the input from IBUU simulations  with $R_{\rm HMT}=15\%$. The solid squares with the grey bands are the output of the RL algorithm. Shown in the inset are the expected spectra in laboratory with the detector response filtering. }
\end{figure}

Fig.~\ref{AlgorithmTestenough} (a) and (b) presents the energy spectra in a high statistics ($N_{\rm evt}=6.25\times 10^{10}$) and a low statistics ($N_{\rm evt}=6.25\times 10^7$) simulation, respectively, for 25 MeV $\rm {^{86}Kr + ^{124}Sn}$  reactions. The results are presented in center of mass (CM) reference frame. It has been converted to the laboratory reference forth and back in order to adapt the detector response filter. The red dashed curve is the original spectrum with $R_{\rm HMT}=15\%$. The open symbols with the error bars shown in the insets are the expected spectrum of the detector output in laboratory. Taking the theoretic spectrum with $R_{\rm HMT}=0\%$ (not shown)  as the initial spectrum in the first iteration $\mathcal{E}^{(0)}$, one can solve reversely the original  spectrum using the RL algorithm with the formulae (9) to (13), as  shown by the solid symbols. Indeed the choice of $\mathcal{E}^{(0)}$ does not affect the final output. For the high statistics test in panel (a), it is shown the reconstructed spectrum (solid symbols) is in agreement with input original one (red dashed curves). To estimate the uncertainty in the gamma spectrum obtained through the RL algorithm, we used multinomial sampling on the central energy spectrum derived from the experimental data. The sampled spectra were individually processed using the RL iterative algorithm, yielding a set of reconstructed spectra. For each bin in these spectra, the standard deviation across the set was calculated and used as the uncertainty in the final extracted spectrum. For the low statistics test in panel (b), the reconstructed data points situates also on top of the original curve, while the statistical uncertainty are significant. It is demonstrated that the RL algorithm is very effective in solving the $\gamma$ original energy spectrum. Clearly,  when the counts of measurements increase, the smoothness of inverse solution will become better. It is then clearly favorable to increase the statistics in order to reduce the error of the inverse solution results and to improve the confidence level of the conclusion in real applications.

\section{Application to Experimental Results Analysis}\label{secanalysis}

\subsection{Experiment Data Solution}

The RL algorithms discussed above are applied to reconstruct the original energy spectrum of the Bremsstrahlung  $\gamma$ emitted in 25 MeV/u \ce{^{86}Kr + ^{124}Sn} performed at CSHINE  \cite{QIN2023168330,QIN2024138514}. 

\begin{figure}[htbp]
\includegraphics[width=0.4\textwidth]{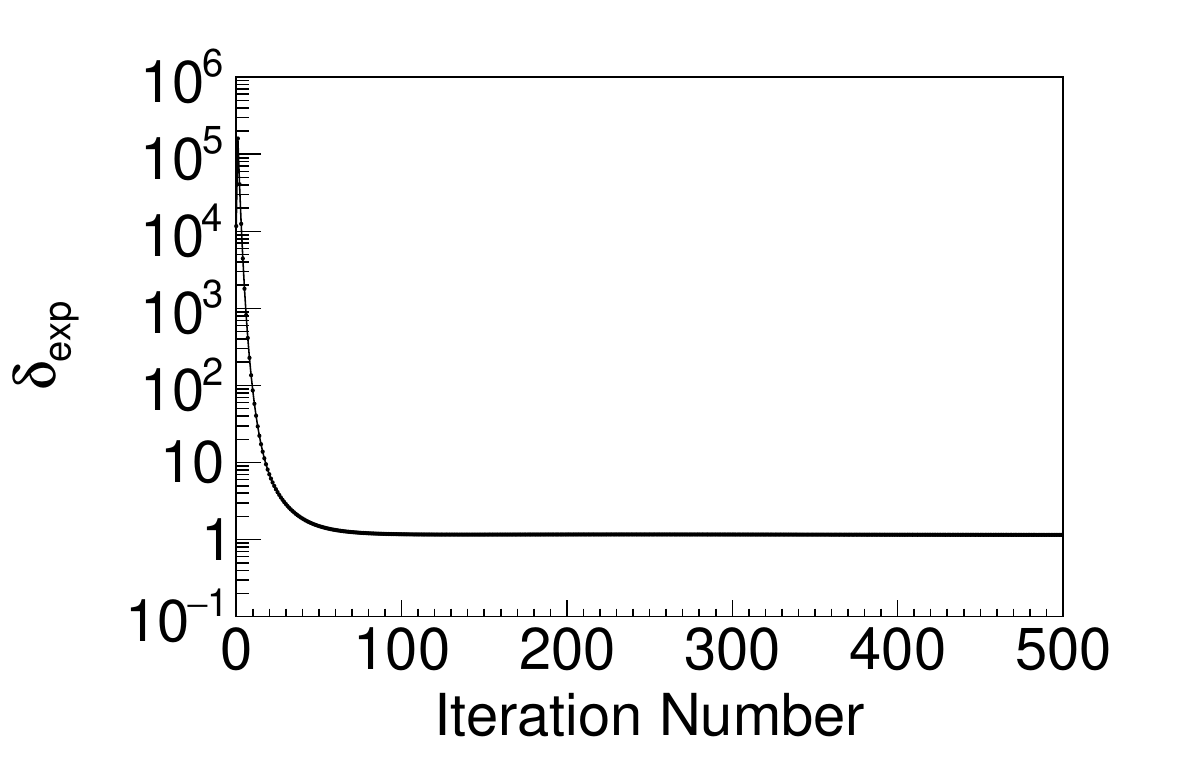}
\caption{\label{stdTest_exp} (Color Online)  The variation of  the standard deviation $\delta_{\rm exp}$ between the experimental spectrum and the predicted spectrum in measurement as a function of iteration time in solving the high-energy $\gamma$ spectrum reversely.}
\end{figure}

\begin{figure*}[htpb]
\includegraphics[width=\textwidth]{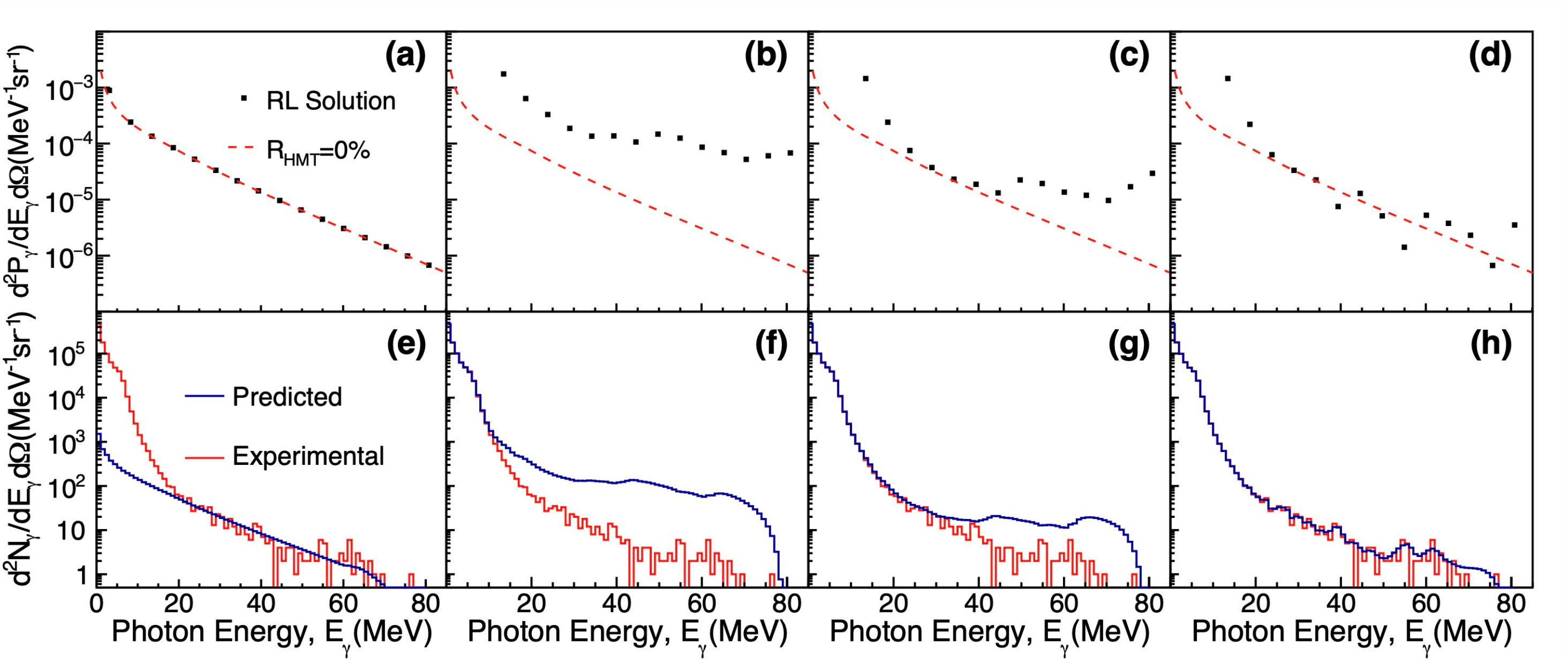}
\caption{\label{IterationRcordexp} (Color Online) The the output of the RL algorithm for the original spectrum (upper row) and the predicted spectrum in measurement (lower row)  after   $0^{\rm th}$ (a,e), $5^{\rm th}$ (b,f), $10^{\rm th}$ (c,g) and $100^{\rm th}$ (d,h) iteration, respectively. See text for details.}
\end{figure*}

Iteration times is an important factor to ensure the convergence, meanwhile to keep the computational efficiency. The terminating point of the iteration is identified by  monitoring  the standard deviation $\delta_{\rm exp}$ of the experimental spectrum and the predicted spectrum in detector response after each iteration. The quantity $\delta_{\rm exp}$ is defined as the following,
\begin{equation}
    \delta_{\rm exp} = \sum_i(\frac{e_i^P-e_i}{\sqrt{e_i}})^2/N_p,
\end{equation}
here, $e_i$ is the input measurement value of the corresponding $i^{\rm th}$ energy point and $e_i^P$ is the predicted value obtained in the current iterations. $N_p$ is the number of bins. If $\sqrt{e_i}$ in the denominator equals zero, it is replaced by $\sqrt{e_i^P}$ to avoid a segment fault. The distribution of $\delta_{\rm exp}$ as function of iteration time is depicted in Fig.~\ref{stdTest_exp}. It is shown that after a certain number of iterations, the value of $\delta_{\rm exp}$ gradually converges and remains constant. The final nonzero  $\delta_{\rm exp}$  value mainly comes from the contribution of the null bins where the experimental statistics is zero.

 In order to gain a more detailed view on the iteration process, we present in Fig.~\ref{IterationRcordexp} the output of the RL algorithm for the original spectrum (upper row) and the predicted spectrum in measurement (lower row)  after the $0^{\rm th}$, $5^{\rm th}$, $10^{\rm th}$ and $100^{\rm th}$ iteration, respectively.  As the initial input, we adapt the theoretic spectrum with $R_{\rm HMT}=0\%$ as the real original spectrum, as shown by the red dashed curve in panel (a). After the detector response filter, one expects the spectrum of the detector output, shown by the blue histogram in panel (e), it exhibits large disagreement with the experimentally measured spectrum shown by the red histogram. After 5 iterations, the low energy part is fast converged because in the low energy part the full energy can easily be collected by the detector and only the diagonal matrix element is nonzero. With the interation times increasing, the predicted spectrum of the detector output approaches gradually to the real one, as shown in panels (f-g). After 100 times of the iteration, as shown in (h) the predicted spectrum of detector output is in good agreement with the experimental spectrum. Meanwhile, as shown in panel (d), the original $\gamma$ spectrum is situating in the vicinity of the IBUU calculation,  except for some jumping  points at high energy side,  originated from the large fluctuations on the experimental spectrum.
 
\begin{figure}[htbp]
\includegraphics[width=0.4\textwidth]{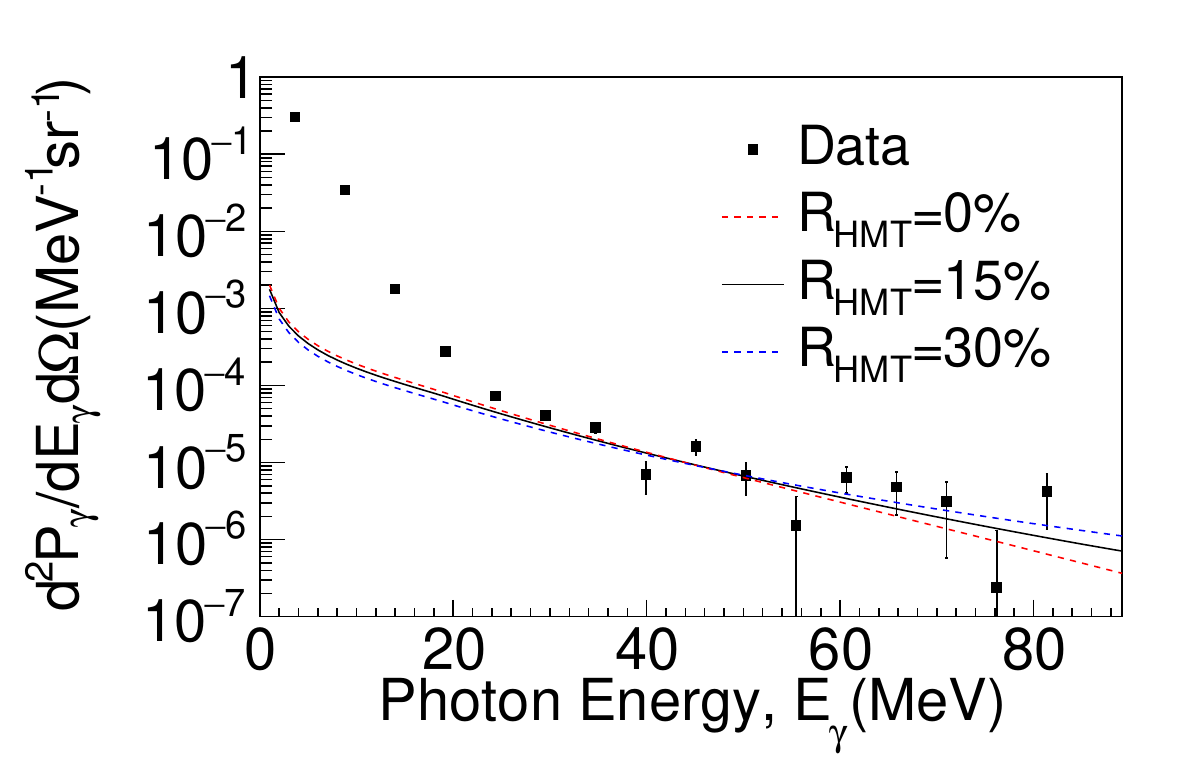}
\caption{\label{ExperimentDirect} (Color Online)  The solution of the original  $\gamma$ spectrum (solid symbols) in comparison to the theoretic predictions from IBUU transport model.}
\end{figure}

Thus, in our analysis, we set conservatively the iteration number to $500$ in RL algorithm implementation. The inverse solution of high-energy $\gamma$ spectrum in 25 MeV/u  \ce{^{86}Kr + ^{124}Sn}  is presented in Fig.~\ref{ExperimentDirect} in comparison to IBUU predictions with $R_{\rm HMT}=0\%$, 15\% and 30\%, respectively. The central values of scatter points are reconstructed  by RL algorithm  from the input measured spectrum and the error bars are derived from the standard deviations  when one samples the content in the corresponding bin using multinomial distribution to ensure that the total count of our pseudo-gamma spectrum matches that of the experimental statistics. Appendix \ref{OrigSpecTable} and \ref{ExpSpecTable} present the reconstructed original spectrum (in CM) using RL method and the experimental spectrum of detector output (in laboratory), respectively. 
With these results, one can readily analyze the most probable HMT ratio of nucleons in nuclei.

\subsection{HMT Analysis}

Next, we are to determine the favored ratio  $R_{\rm HMT}$ from the RL solution. Let's define $\chi^2_c$ to evaluate the closeness (overall deviation) of the theoretic spectrum to the inversely solved original $\gamma$ spectrum in the experiment,
\begin{equation}
\chi^2_c = \frac{1}{n_{\rm DOF}}\sum_i\frac{1}{\sigma^2_{i}}\left[\left(\frac{d^2P_{\gamma}}{dE_{\gamma}d\Omega}\right)_{i}-\left(\frac{d^2P^{(c)}_{\gamma}}{dE_{\gamma}d\Omega}\right)_{i}\right]^2,
\end{equation}
where the two terms in the square bracket denote the possibility in the $i^{\rm th}$ bin of energy for experiment and theoretic prediction, respectively. The super and subscript $c$ represents  the percentage of HMT in IBUU simulations, $\sigma_i$ in the denominator is the standard deviation obtained from the sampling simulation and $n_{\rm DOF}$ refers to the number of degree of freedom (DOF), which is approximately taken as the number of data points in the corresponding energy range. In addition, the scales of different theoretical curves shall be adjusted to minimize the $\chi^2_c$ (the $\chi^2_c$s mentioned in the following process are all minimized). This quantity is used to evaluate the closeness  of the inversely solved energy spectrum to various spectra of IBUU model simulations.

\begin{table}[htbp]
\caption{\label{Chi2of30to75} $\chi^2_c$s of direct solution high-energy $\gamma$ spectrum compared with different HMT percentages in the energy ROI from 35 to 80 MeV.}
\begin{ruledtabular}
\begin{tabular}{cccc}
$R_{\rm HMT}$ &$\chi^2_c$ &$R_{\rm HMT}$ &$\chi^2_c$\\
\hline
$0\%$ & 1.936717 & $20\%$ & 1.839244 \\
$5\%$ & 1.900685 & $25\%$ & 1.835009 \\
$10\%$ & 1.871585 & $30\%$ & 1.839868 \\
$15\%$ & 1.851869 & &\\ 
\end{tabular}
\end{ruledtabular}
\end{table}

Energies below $E_{\gamma}< 30$ $\rm MeV$ are largely influenced by collective resonance and statistical emissions, which are excluded from this analysis. At energies higher than 80 MeV, the impact of cosmic rays becomes significant, limiting the applicability of the IBUU model. Table.~\ref{Chi2of30to75} lists the $\chi^2_c$ values using different $R_{\rm HMT}$ in the energy range of interest (ROI) from 35 to 80 MeV. From the table,  the minimum $\chi^2_c$ value is found at $R_{\rm HMT}=25\%$. It indicates that the HMT exists in nuclei, qualitatively consistent with our previous analysis. On the other hand, the most favored ratio of  $R_{\rm HMT}=25\%$ is somewhat high than the most probable value derived in \cite{QIN2024138514}, see next.

It is also necessary to conduct the hypothesis test in order to determine the probability to exclude the  $R_{\rm HMT}=0\%$, i.e., to exclude the assumption of null high momentum tail of nucleons in nuclei. To achieve this goal, the theoretic curve of  $R_{\rm HMT}=0\%$ is multiplied by a factor $\eta N_{\rm evt}$ to have a hypothetical spectrum with certain counts in each bin without fluctuation. Here $N_{\rm evt}=51269108$ is the number of events recorded in experiment, and $\eta=1.22$ is a normalization factor to make the summing counts above 20 MeV equal to that of experimental data.  With this hypothetical spectrum, one can generate a psudo-experimental spectrum, taking the fluctuations into account  by sampling the content in each bin of $E_{\gamma}$ with multinomial distribution. Starting with the  psudo-experimental spectra, one can repeatedly use the detector response filter and solve the  RL algorithm to get the original $\gamma$ spectra with the zero HMT assumptions. By computing the difference of $\chi^2_{0}-\chi^2_{25}$ with (16) for the psudo-experimental spectra in the same way as done in the real experiment, one can infer the possibility to conclude a non-zero HMT ($R_{\rm HMT}=25\%$ here)  where the input condition is $R_{\rm HMT}=0\%$.

\begin{figure}[htbp]
\includegraphics[width=0.4\textwidth]{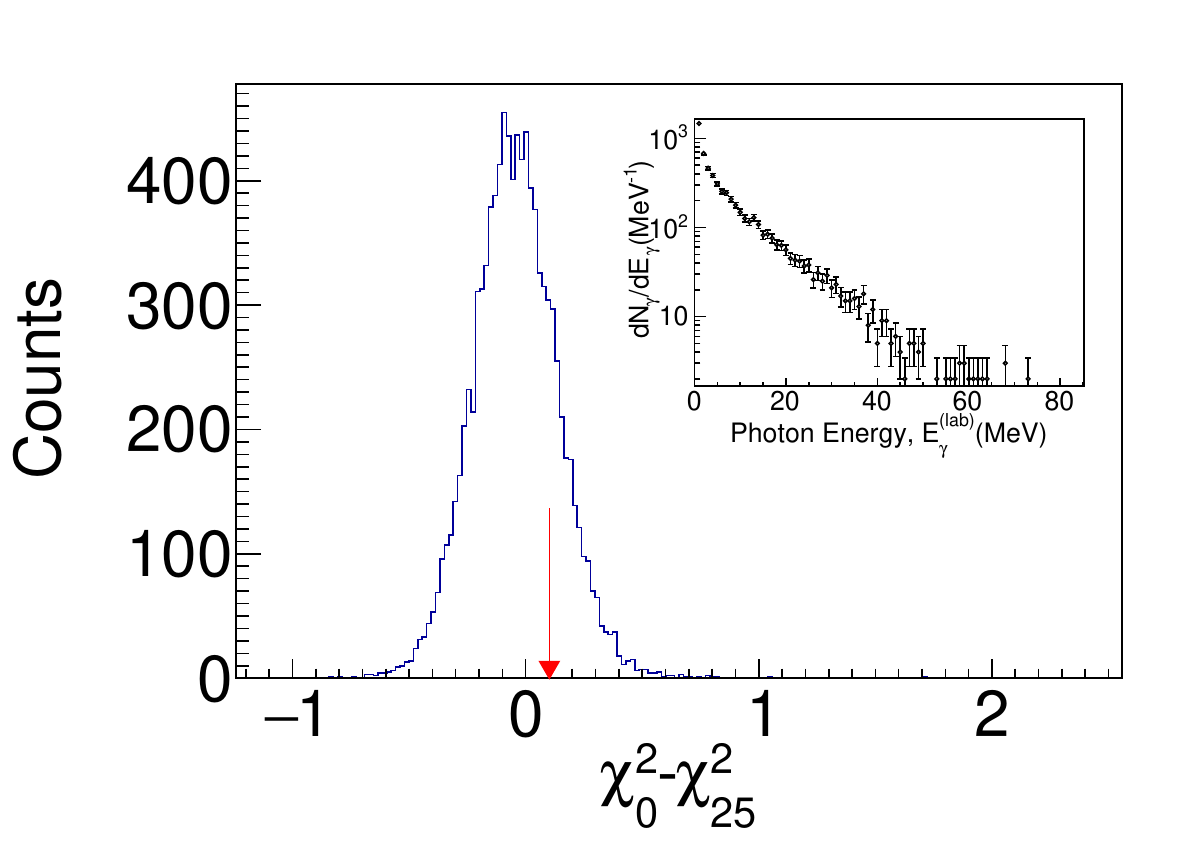}
\caption{\label{DChi2HMT0} (Color Online) Distribution of $\chi^2_{0}-\chi^2_{25}$ for 10000 sampled psudo-experimental spectra in the energy ROI from 35 to  80 MeV. The red arrow indicates the value of the experiment. The inset displays one example of the predicted spectra as detector output in laboratory.}
\end{figure}

Fig.~\ref{DChi2HMT0} presents the histogram of $\chi^2_{0}-\chi^2_{25}$ for 10000 sampled psudo-experimental spectra sampled with  $R_{\rm HMT}=0\%$. The inset displays one example of the spectra of the detector output in laboratory frame. The energy ROI is similarly taken from 35 to 80 MeV. The red arrow at 0.102 indicates the position of the experimental result of $\chi^2_{0}-\chi^2_{25}$, the area on the left (right) side of the arrow adds up to 80\% (20\%), suggesting that there is a probability of 20\% to observe a non-zero HMT behavior from a "gedanken experiment" without HMT on the nucleon momentum distribution. Equivalently, it is stated that the experimental results exclude the zero HMT hypothesis  at 80\%  confidential level.

\begin{table*}[htbp]
\caption{\label{MoreEnergyRanges}  The $\chi^2_{c}$s between experimental solution results and theoretical models at different HMT ratios in different energy ROI, the HMT ratio preferences, the confidence level of supporting the favored $R_{\rm HMT}$ are also listed.}
\begin{ruledtabular}
\begin{tabular}{c|cccccc}
$E_{\gamma}$ Ranges ($\rm MeV$) &30-70 &30-75 &30-80 &35-70 &35-75 &35-80\\
\hline
$R_{\rm HMT}$ &\multicolumn{6}{c}{$\chi_c^2$}\\
\hline
$0\%$&2.677 &2.353 &2.125 &2.539 &2.207 &1.937 \\
$5\%$&2.669 &2.335 &2.137 &2.475 &2.141 &1.901  \\
$10\%$&2.674 &2.327 &2.167 &2.413 &2.077 &1.872   \\
$15\%$&2.681 &2.324 &2.206 &2.359 &2.021 &1.852   \\
$20\%$&2.702 &2.333 &2.270 &2.300 &1.960 &1.839   \\
$25\%$&2.714 &2.340 &2.297 &2.279 &1.938 &1.835    \\
$30\%$&2.788 &2.394 &2.454 &2.199 &1.857 &1.840  \\
\hline
Favored $R_{\rm HMT}$ &$5\%$ &$15\%$ &$0\%$ &$\ge 30\%$\footnote{\label{foot1}Beyond $30\%$, the $\chi^2_c$ in these two columns decreases with  $R_{\rm HMT}$ due to the existence of the last data points, but the difference of the hardness of the spectrum by varying $R_{\rm HMT}$  becomes less pronounced and smaller than the uncertainty. Thus the calculations is terminated at $30\%$.} &$\ge 30\%$\footref{foot1} &$25\%$  \\
\hline
C.L. (\%) &43 &57 &- &93 &96 &80 \\
\end{tabular}
\end{ruledtabular}
\end{table*}

Finally, we check the look-elsewhere effect. Tab.~\ref{MoreEnergyRanges} lists the values of $\chi^2_{c}$ at different $R_{\rm HMT}$ and in different energy ROI, together with the favored $R_{\rm HMT}$ and the corresponding confidential level to exclude the zero HMT assumption. It is shown that the favored ratio of  $R_{\rm HMT}$ in different analysis ranges is changed. With the only exception in ROI from 30 to 80 MeV, all the other analyses favor a non zero HMT, with the best value depends on the ROI. The confidential level varies from 42\% to 96\%, and the average  confidential level of excluding the zero HMT hypothesis is about $74\%$,  suggesting the existence of SRC effect in nuclei.

The results from this new analysis is in qualitative agreement with our previous conclusion by analyzing directly the measured spectrum  without introducing the RL method to solve the inverse problem \cite{QIN2024138514}. Considering the small difference of  $\chi^2_{c}$ here as well as the small difference of the likelihood by varying $R_{\rm HMT}$ in \cite{QIN2024138514}, it is understandable that the favored value of $R_{\rm HMT}$ exhibits some variation. Meanwhile, due to the low experimental statistics,  the confidential level to exclude zero  $R_{\rm HMT}$ is yet not high, i.e., in the vicinity of $2\sigma$ at most. In order to increase the resolution power of HMT ratio from the Bremsstrahlung $\gamma$ emission, it is necessary to increase the statistics and to extend  the high energy border because the discrepancy caused by different  $R_{\rm HMT}$ is increasingly pronounced if one goes to higher $\gamma$ energy. Further efforts are accordingly required in the future experiments \cite{Si:2023dmz}.

\section{Conclusions}

For the first time, the Richardson-Lucy  algorithm has been applied to solve inversely the original  energy spectrum of Bremsstrahlung $\gamma$-rays in heavy ion reactions. The $\gamma$ spectrum in 25 MeV/u $^{86}$Kr+$^{124}$Sn measured at CSHINE are reconstructed and compared to the isospin- and momentum-dependent Boltzmann-Uehling-Uhlenbeck transport model simulations. It is  inferred that a ratio of high momentum tail $R_{\rm HMT}=25\%$ describes the experimental data points with the least $\chi^2_c$ in the ROI from 35 to 80 MeV and the confidential level to exclude the zero  $R_{\rm HMT}$ is averagely 74\%, in qualitative agreement with our previous conclusion without introducing the RL inverse solution process. 

Although the application of the RL method has its advantage that the detector response  has already been folded in solving the inverse problem to reconstruct the original $\gamma$ spectrum, the resolution power to quantify the HMT ratio  is indeed determined by the statistics and the upper limit of the $\gamma$ energy covered by the experiment. So, to explore further the SRC effect in atomic nuclei with enhanced accuracy from the Bremsstrahlung $\gamma$ emission in heavy ion reactions, further experimental efforts are called for. 

{\it  Acknowledgements.}
This work is supported  by the Ministry of Science and Technology of China under Grant Nos. 2022YFE0103400,
 by the National Natural Science Foundation of China under Grant Nos. 12335008 and 12275129  
 and by Tsinghua University Initiative Scientific Research Program.

\bibliography{reference}

\begin{thebibliography}{27}%
\makeatletter
\providecommand \@ifxundefined [1]{%
 \@ifx{#1\undefined}
}%
\providecommand \@ifnum [1]{%
 \ifnum #1\expandafter \@firstoftwo
 \else \expandafter \@secondoftwo
 \fi
}%
\providecommand \@ifx [1]{%
 \ifx #1\expandafter \@firstoftwo
 \else \expandafter \@secondoftwo
 \fi
}%
\providecommand \natexlab [1]{#1}%
\providecommand \enquote  [1]{``#1''}%
\providecommand \bibnamefont  [1]{#1}%
\providecommand \bibfnamefont [1]{#1}%
\providecommand \citenamefont [1]{#1}%
\providecommand \href@noop [0]{\@secondoftwo}%
\providecommand \href [0]{\begingroup \@sanitize@url \@href}%
\providecommand \@href[1]{\@@startlink{#1}\@@href}%
\providecommand \@@href[1]{\endgroup#1\@@endlink}%
\providecommand \@sanitize@url [0]{\catcode `\\12\catcode `\$12\catcode `\&12\catcode `\#12\catcode `\^12\catcode `\_12\catcode `\%12\relax}%
\providecommand \@@startlink[1]{}%
\providecommand \@@endlink[0]{}%
\providecommand \url  [0]{\begingroup\@sanitize@url \@url }%
\providecommand \@url [1]{\endgroup\@href {#1}{\urlprefix }}%
\providecommand \urlprefix  [0]{URL }%
\providecommand \Eprint [0]{\href }%
\providecommand \doibase [0]{https://doi.org/}%
\providecommand \selectlanguage [0]{\@gobble}%
\providecommand \bibinfo  [0]{\@secondoftwo}%
\providecommand \bibfield  [0]{\@secondoftwo}%
\providecommand \translation [1]{[#1]}%
\providecommand \BibitemOpen [0]{}%
\providecommand \bibitemStop [0]{}%
\providecommand \bibitemNoStop [0]{.\EOS\space}%
\providecommand \EOS [0]{\spacefactor3000\relax}%
\providecommand \BibitemShut  [1]{\csname bibitem#1\endcsname}%
\let\auto@bib@innerbib\@empty
\bibitem [{\citenamefont {Hen}\ \emph {et~al.}(2017)\citenamefont {Hen}, \citenamefont {Miller}, \citenamefont {Piasetzky},\ and\ \citenamefont {Weinstein}}]{RevModPhys.89.045002}%
  \BibitemOpen
  \bibfield  {author} {\bibinfo {author} {\bibfnamefont {O.}~\bibnamefont {Hen}}, \bibinfo {author} {\bibfnamefont {G.~A.}\ \bibnamefont {Miller}}, \bibinfo {author} {\bibfnamefont {E.}~\bibnamefont {Piasetzky}},\ and\ \bibinfo {author} {\bibfnamefont {L.~B.}\ \bibnamefont {Weinstein}},\ }\bibfield  {title} {\bibinfo {title} {Nucleon-nucleon correlations, short-lived excitations, and the quarks within},\ }\href {https://doi.org/10.1103/RevModPhys.89.045002} {\bibfield  {journal} {\bibinfo  {journal} {Rev. Mod. Phys.}\ }\textbf {\bibinfo {volume} {89}},\ \bibinfo {pages} {045002} (\bibinfo {year} {2017})}\BibitemShut {NoStop}%
\bibitem [{\citenamefont {Aubert}\ \emph {et~al.}(1983)\citenamefont {Aubert} \emph {et~al.}}]{EuropeanMuon:1983wih}%
  \BibitemOpen
  \bibfield  {author} {\bibinfo {author} {\bibfnamefont {J.~J.}\ \bibnamefont {Aubert}} \emph {et~al.} (\bibinfo {collaboration} {European Muon}),\ }\bibfield  {title} {\bibinfo {title} {{The ratio of the nucleon structure functions $F2_n$ for iron and deuterium}},\ }\href {https://doi.org/10.1016/0370-2693(83)90437-9} {\bibfield  {journal} {\bibinfo  {journal} {Phys. Lett. B}\ }\textbf {\bibinfo {volume} {123}},\ \bibinfo {pages} {275} (\bibinfo {year} {1983})}\BibitemShut {NoStop}%
\bibitem [{\citenamefont {Hen}\ \emph {et~al.}(2014)\citenamefont {Hen} \emph {et~al.}}]{Hen:2014nza}%
  \BibitemOpen
  \bibfield  {author} {\bibinfo {author} {\bibfnamefont {O.}~\bibnamefont {Hen}} \emph {et~al.},\ }\bibfield  {title} {\bibinfo {title} {{Momentum sharing in imbalanced Fermi systems}},\ }\href {https://doi.org/10.1126/science.1256785} {\bibfield  {journal} {\bibinfo  {journal} {Science}\ }\textbf {\bibinfo {volume} {346}},\ \bibinfo {pages} {614} (\bibinfo {year} {2014})},\ \Eprint {https://arxiv.org/abs/1412.0138} {arXiv:1412.0138 [nucl-ex]} \BibitemShut {NoStop}%
\bibitem [{\citenamefont {Duer}\ \emph {et~al.}(2018)\citenamefont {Duer} \emph {et~al.}}]{CLAS:2018yvt}%
  \BibitemOpen
  \bibfield  {author} {\bibinfo {author} {\bibfnamefont {M.}~\bibnamefont {Duer}} \emph {et~al.} (\bibinfo {collaboration} {CLAS}),\ }\bibfield  {title} {\bibinfo {title} {{Probing high-momentum protons and neutrons in neutron-rich nuclei}},\ }\href {https://doi.org/10.1038/s41586-018-0400-z} {\bibfield  {journal} {\bibinfo  {journal} {Nature}\ }\textbf {\bibinfo {volume} {560}},\ \bibinfo {pages} {617} (\bibinfo {year} {2018})}\BibitemShut {NoStop}%
\bibitem [{\citenamefont {Schmookler}\ \emph {et~al.}(2019)\citenamefont {Schmookler} \emph {et~al.}}]{CLAS:2019vsb}%
  \BibitemOpen
  \bibfield  {author} {\bibinfo {author} {\bibfnamefont {B.}~\bibnamefont {Schmookler}} \emph {et~al.} (\bibinfo {collaboration} {CLAS}),\ }\bibfield  {title} {\bibinfo {title} {{Modified structure of protons and neutrons in correlated pairs}},\ }\href {https://doi.org/10.1038/s41586-019-0925-9} {\bibfield  {journal} {\bibinfo  {journal} {Nature}\ }\textbf {\bibinfo {volume} {566}},\ \bibinfo {pages} {354} (\bibinfo {year} {2019})},\ \Eprint {https://arxiv.org/abs/2004.12065} {arXiv:2004.12065 [nucl-ex]} \BibitemShut {NoStop}%
\bibitem [{\citenamefont {Li}\ \emph {et~al.}(2022)\citenamefont {Li} \emph {et~al.}}]{Li:2022fhh}%
  \BibitemOpen
  \bibfield  {author} {\bibinfo {author} {\bibfnamefont {S.}~\bibnamefont {Li}} \emph {et~al.},\ }\bibfield  {title} {\bibinfo {title} {{Revealing the short-range structure of the mirror nuclei $^{3}$H and $^{3}$He}},\ }\href {https://doi.org/10.1038/s41586-022-05007-2} {\bibfield  {journal} {\bibinfo  {journal} {Nature}\ }\textbf {\bibinfo {volume} {609}},\ \bibinfo {pages} {41} (\bibinfo {year} {2022})},\ \Eprint {https://arxiv.org/abs/2210.04189} {arXiv:2210.04189 [nucl-ex]} \BibitemShut {NoStop}%
\bibitem [{\citenamefont {Tang}\ \emph {et~al.}(2003)\citenamefont {Tang} \emph {et~al.}}]{Tang:2002ww}%
  \BibitemOpen
  \bibfield  {author} {\bibinfo {author} {\bibfnamefont {A.}~\bibnamefont {Tang}} \emph {et~al.},\ }\bibfield  {title} {\bibinfo {title} {{n-p short range correlations from (p,2p + n) measurements}},\ }\href {https://doi.org/10.1103/PhysRevLett.90.042301} {\bibfield  {journal} {\bibinfo  {journal} {Phys. Rev. Lett.}\ }\textbf {\bibinfo {volume} {90}},\ \bibinfo {pages} {042301} (\bibinfo {year} {2003})},\ \Eprint {https://arxiv.org/abs/nucl-ex/0206003} {arXiv:nucl-ex/0206003} \BibitemShut {NoStop}%
\bibitem [{\citenamefont {Piasetzky}\ \emph {et~al.}(2006)\citenamefont {Piasetzky}, \citenamefont {Sargsian}, \citenamefont {Frankfurt}, \citenamefont {Strikman},\ and\ \citenamefont {Watson}}]{Piasetzky:2006ai}%
  \BibitemOpen
  \bibfield  {author} {\bibinfo {author} {\bibfnamefont {E.}~\bibnamefont {Piasetzky}}, \bibinfo {author} {\bibfnamefont {M.}~\bibnamefont {Sargsian}}, \bibinfo {author} {\bibfnamefont {L.}~\bibnamefont {Frankfurt}}, \bibinfo {author} {\bibfnamefont {M.}~\bibnamefont {Strikman}},\ and\ \bibinfo {author} {\bibfnamefont {J.~W.}\ \bibnamefont {Watson}},\ }\bibfield  {title} {\bibinfo {title} {{Evidence for the strong dominance of proton-neutron correlations in nuclei}},\ }\href {https://doi.org/10.1103/PhysRevLett.97.162504} {\bibfield  {journal} {\bibinfo  {journal} {Phys. Rev. Lett.}\ }\textbf {\bibinfo {volume} {97}},\ \bibinfo {pages} {162504} (\bibinfo {year} {2006})},\ \Eprint {https://arxiv.org/abs/nucl-th/0604012} {arXiv:nucl-th/0604012} \BibitemShut {NoStop}%
\bibitem [{\citenamefont {Xue}\ \emph {et~al.}(2016)\citenamefont {Xue}, \citenamefont {Xu}, \citenamefont {Yong},\ and\ \citenamefont {Ren}}]{XUE2016486}%
  \BibitemOpen
  \bibfield  {author} {\bibinfo {author} {\bibfnamefont {H.}~\bibnamefont {Xue}}, \bibinfo {author} {\bibfnamefont {C.}~\bibnamefont {Xu}}, \bibinfo {author} {\bibfnamefont {G.-C.}\ \bibnamefont {Yong}},\ and\ \bibinfo {author} {\bibfnamefont {Z.}~\bibnamefont {Ren}},\ }\bibfield  {title} {\bibinfo {title} {Neutron--proton bremsstrahlung as a possible probe of high-momentum component in nucleon momentum distribution},\ }\href {https://doi.org/https://doi.org/10.1016/j.physletb.2016.02.044} {\bibfield  {journal} {\bibinfo  {journal} {Physics Letters B}\ }\textbf {\bibinfo {volume} {755}},\ \bibinfo {pages} {486} (\bibinfo {year} {2016})}\BibitemShut {NoStop}%
\bibitem [{\citenamefont {Qin}\ \emph {et~al.}(2024)\citenamefont {Qin}, \citenamefont {Niu}, \citenamefont {Guo}, \citenamefont {Xiao}, \citenamefont {Tian}, \citenamefont {Wang}, \citenamefont {Qin}, \citenamefont {Diao}, \citenamefont {Guan}, \citenamefont {Si}, \citenamefont {Zhang}, \citenamefont {Zhang}, \citenamefont {Wei}, \citenamefont {Yang}, \citenamefont {Ma}, \citenamefont {Zou}, \citenamefont {Qiu}, \citenamefont {Huang}, \citenamefont {Hu}, \citenamefont {Duan}, \citenamefont {Ong}, \citenamefont {Yang}, \citenamefont {Xu}, \citenamefont {Wang}, \citenamefont {Bai}, \citenamefont {Ma}, \citenamefont {Duan}, \citenamefont {Yang}, \citenamefont {Hu}, \citenamefont {Wang}, \citenamefont {Sun}, \citenamefont {Maydanyuk}, \citenamefont {Xu},\ and\ \citenamefont {Xiao}}]{QIN2024138514}%
  \BibitemOpen
  \bibfield  {author} {\bibinfo {author} {\bibfnamefont {Y.}~\bibnamefont {Qin}}, \bibinfo {author} {\bibfnamefont {Q.}~\bibnamefont {Niu}}, \bibinfo {author} {\bibfnamefont {D.}~\bibnamefont {Guo}}, \bibinfo {author} {\bibfnamefont {S.}~\bibnamefont {Xiao}}, \bibinfo {author} {\bibfnamefont {B.}~\bibnamefont {Tian}}, \bibinfo {author} {\bibfnamefont {Y.}~\bibnamefont {Wang}}, \bibinfo {author} {\bibfnamefont {Z.}~\bibnamefont {Qin}}, \bibinfo {author} {\bibfnamefont {X.}~\bibnamefont {Diao}}, \bibinfo {author} {\bibfnamefont {F.}~\bibnamefont {Guan}}, \bibinfo {author} {\bibfnamefont {D.}~\bibnamefont {Si}}, \bibinfo {author} {\bibfnamefont {B.}~\bibnamefont {Zhang}}, \bibinfo {author} {\bibfnamefont {Y.}~\bibnamefont {Zhang}}, \bibinfo {author} {\bibfnamefont {X.}~\bibnamefont {Wei}}, \bibinfo {author} {\bibfnamefont {H.}~\bibnamefont {Yang}}, \bibinfo {author} {\bibfnamefont {P.}~\bibnamefont {Ma}}, \bibinfo {author} {\bibfnamefont {H.}~\bibnamefont {Zou}}, \bibinfo {author} {\bibfnamefont
  {T.}~\bibnamefont {Qiu}}, \bibinfo {author} {\bibfnamefont {X.}~\bibnamefont {Huang}}, \bibinfo {author} {\bibfnamefont {R.}~\bibnamefont {Hu}}, \bibinfo {author} {\bibfnamefont {L.}~\bibnamefont {Duan}}, \bibinfo {author} {\bibfnamefont {H.~J.}\ \bibnamefont {Ong}}, \bibinfo {author} {\bibfnamefont {Y.}~\bibnamefont {Yang}}, \bibinfo {author} {\bibfnamefont {S.}~\bibnamefont {Xu}}, \bibinfo {author} {\bibfnamefont {K.}~\bibnamefont {Wang}}, \bibinfo {author} {\bibfnamefont {Z.}~\bibnamefont {Bai}}, \bibinfo {author} {\bibfnamefont {J.}~\bibnamefont {Ma}}, \bibinfo {author} {\bibfnamefont {F.}~\bibnamefont {Duan}}, \bibinfo {author} {\bibfnamefont {G.}~\bibnamefont {Yang}}, \bibinfo {author} {\bibfnamefont {Q.}~\bibnamefont {Hu}}, \bibinfo {author} {\bibfnamefont {H.}~\bibnamefont {Wang}}, \bibinfo {author} {\bibfnamefont {B.}~\bibnamefont {Sun}}, \bibinfo {author} {\bibfnamefont {S.~P.}\ \bibnamefont {Maydanyuk}}, \bibinfo {author} {\bibfnamefont {C.}~\bibnamefont {Xu}},\ and\ \bibinfo {author}
  {\bibfnamefont {Z.}~\bibnamefont {Xiao}},\ }\bibfield  {title} {\bibinfo {title} {Probing high-momentum component in nucleon momentum distribution by neutron-proton bremsstrahlung γ-rays in heavy ion reactions},\ }\href {https://doi.org/https://doi.org/10.1016/j.physletb.2024.138514} {\bibfield  {journal} {\bibinfo  {journal} {Physics Letters B}\ }\textbf {\bibinfo {volume} {850}},\ \bibinfo {pages} {138514} (\bibinfo {year} {2024})}\BibitemShut {NoStop}%
\bibitem [{\citenamefont {Richardson}(1972)}]{Richardson:72}%
  \BibitemOpen
  \bibfield  {author} {\bibinfo {author} {\bibfnamefont {W.~H.}\ \bibnamefont {Richardson}},\ }\bibfield  {title} {\bibinfo {title} {Bayesian-based iterative method of image restoration$\ast$},\ }\href {https://doi.org/10.1364/JOSA.62.000055} {\bibfield  {journal} {\bibinfo  {journal} {J. Opt. Soc. Am.}\ }\textbf {\bibinfo {volume} {62}},\ \bibinfo {pages} {55} (\bibinfo {year} {1972})}\BibitemShut {NoStop}%
\bibitem [{\citenamefont {{Lucy}}(1974)}]{1974AJ.....79..745L}%
  \BibitemOpen
  \bibfield  {author} {\bibinfo {author} {\bibfnamefont {L.~B.}\ \bibnamefont {{Lucy}}},\ }\bibfield  {title} {\bibinfo {title} {{An iterative technique for the rectification of observed distributions}},\ }\href {https://doi.org/10.1086/111605} {\bibfield  {journal} {\bibinfo  {journal} {Astronomical Journal}\ }\textbf {\bibinfo {volume} {79}},\ \bibinfo {pages} {745} (\bibinfo {year} {1974})}\BibitemShut {NoStop}%
\bibitem [{\citenamefont {Fagun~Vankawala}(2015)}]{10.5120/20396-2697}%
  \BibitemOpen
  \bibfield  {author} {\bibinfo {author} {\bibfnamefont {A.~P.}\ \bibnamefont {Fagun~Vankawala}, \bibfnamefont {Amit~Ganatra}},\ }\bibfield  {title} {\bibinfo {title} {A survey on different image deblurring techniques},\ }\href {https://doi.org/10.5120/20396-2697} {\bibfield  {journal} {\bibinfo  {journal} {International Journal of Computer Applications}\ }\textbf {\bibinfo {volume} {116}},\ \bibinfo {pages} {15} (\bibinfo {year} {2015})}\BibitemShut {NoStop}%
\bibitem [{\citenamefont {Danielewicz}\ and\ \citenamefont {Kurata-Nishimura}(2022)}]{PhysRevC.105.034608}%
  \BibitemOpen
  \bibfield  {author} {\bibinfo {author} {\bibfnamefont {P.}~\bibnamefont {Danielewicz}}\ and\ \bibinfo {author} {\bibfnamefont {M.}~\bibnamefont {Kurata-Nishimura}},\ }\bibfield  {title} {\bibinfo {title} {Deblurring for nuclei: 3d characteristics of heavy-ion collisions},\ }\href {https://doi.org/10.1103/PhysRevC.105.034608} {\bibfield  {journal} {\bibinfo  {journal} {Phys. Rev. C}\ }\textbf {\bibinfo {volume} {105}},\ \bibinfo {pages} {034608} (\bibinfo {year} {2022})}\BibitemShut {NoStop}%
\bibitem [{\citenamefont {Nzabahimana}\ and\ \citenamefont {Danielewicz}(2023)}]{NZABAHIMANA2023138247}%
  \BibitemOpen
  \bibfield  {author} {\bibinfo {author} {\bibfnamefont {P.}~\bibnamefont {Nzabahimana}}\ and\ \bibinfo {author} {\bibfnamefont {P.}~\bibnamefont {Danielewicz}},\ }\bibfield  {title} {\bibinfo {title} {Source function from two-particle correlation through deblurring},\ }\href {https://doi.org/https://doi.org/10.1016/j.physletb.2023.138247} {\bibfield  {journal} {\bibinfo  {journal} {Physics Letters B}\ }\textbf {\bibinfo {volume} {846}},\ \bibinfo {pages} {138247} (\bibinfo {year} {2023})}\BibitemShut {NoStop}%
\bibitem [{\citenamefont {Nzabahimana}\ \emph {et~al.}(2023)\citenamefont {Nzabahimana}, \citenamefont {Redpath}, \citenamefont {Baumann}, \citenamefont {Danielewicz}, \citenamefont {Giuliani},\ and\ \citenamefont {Gu\`eye}}]{PhysRevC.107.064315}%
  \BibitemOpen
  \bibfield  {author} {\bibinfo {author} {\bibfnamefont {P.}~\bibnamefont {Nzabahimana}}, \bibinfo {author} {\bibfnamefont {T.}~\bibnamefont {Redpath}}, \bibinfo {author} {\bibfnamefont {T.}~\bibnamefont {Baumann}}, \bibinfo {author} {\bibfnamefont {P.}~\bibnamefont {Danielewicz}}, \bibinfo {author} {\bibfnamefont {P.}~\bibnamefont {Giuliani}},\ and\ \bibinfo {author} {\bibfnamefont {P.}~\bibnamefont {Gu\`eye}},\ }\bibfield  {title} {\bibinfo {title} {Deconvoluting experimental decay energy spectra: The $^{26}\mathrm{O}$ case},\ }\href {https://doi.org/10.1103/PhysRevC.107.064315} {\bibfield  {journal} {\bibinfo  {journal} {Phys. Rev. C}\ }\textbf {\bibinfo {volume} {107}},\ \bibinfo {pages} {064315} (\bibinfo {year} {2023})}\BibitemShut {NoStop}%
\bibitem [{\citenamefont {Vargas}\ \emph {et~al.}(2013)\citenamefont {Vargas}, \citenamefont {Benlliure},\ and\ \citenamefont {Caama{\~n}o}}]{VARGAS201316}%
  \BibitemOpen
  \bibfield  {author} {\bibinfo {author} {\bibfnamefont {J.}~\bibnamefont {Vargas}}, \bibinfo {author} {\bibfnamefont {J.}~\bibnamefont {Benlliure}},\ and\ \bibinfo {author} {\bibfnamefont {M.}~\bibnamefont {Caama{\~n}o}},\ }\bibfield  {title} {\bibinfo {title} {Unfolding the response of a zero-degree magnetic spectrometer from measurements of the Δ resonance},\ }\href {https://doi.org/https://doi.org/10.1016/j.nima.2012.12.087} {\bibfield  {journal} {\bibinfo  {journal} {Nuclear Instruments and Methods in Physics Research Section A: Accelerators, Spectrometers, Detectors and Associated Equipment}\ }\textbf {\bibinfo {volume} {707}},\ \bibinfo {pages} {16} (\bibinfo {year} {2013})}\BibitemShut {NoStop}%
\bibitem [{\citenamefont {Agostinelli}\ \emph {et~al.}(2003)\citenamefont {Agostinelli} \emph {et~al.}}]{Geant4}%
  \BibitemOpen
  \bibfield  {author} {\bibinfo {author} {\bibfnamefont {S.}~\bibnamefont {Agostinelli}} \emph {et~al.} (\bibinfo {collaboration} {GEANT4}),\ }\bibfield  {title} {\bibinfo {title} {{GEANT4--a simulation toolkit}},\ }\href {https://doi.org/10.1016/S0168-9002(03)01368-8} {\bibfield  {journal} {\bibinfo  {journal} {Nucl. Instrum. Meth. A}\ }\textbf {\bibinfo {volume} {506}},\ \bibinfo {pages} {250} (\bibinfo {year} {2003})}\BibitemShut {NoStop}%
\bibitem [{\citenamefont {Qin}\ \emph {et~al.}(2023)\citenamefont {Qin}, \citenamefont {Guo}, \citenamefont {Xiao}, \citenamefont {Wang}, \citenamefont {Guan}, \citenamefont {Diao}, \citenamefont {Qin}, \citenamefont {Si}, \citenamefont {Zhang}, \citenamefont {Zhang}, \citenamefont {Sun}, \citenamefont {Wei}, \citenamefont {Yang}, \citenamefont {Ma}, \citenamefont {Zou}, \citenamefont {Qiu}, \citenamefont {Huang}, \citenamefont {Hu}, \citenamefont {Duan}, \citenamefont {Duan}, \citenamefont {Hu}, \citenamefont {Ma}, \citenamefont {Xu}, \citenamefont {Bai}, \citenamefont {Yang},\ and\ \citenamefont {Xiao}}]{QIN2023168330}%
  \BibitemOpen
  \bibfield  {author} {\bibinfo {author} {\bibfnamefont {Y.}~\bibnamefont {Qin}}, \bibinfo {author} {\bibfnamefont {D.}~\bibnamefont {Guo}}, \bibinfo {author} {\bibfnamefont {S.}~\bibnamefont {Xiao}}, \bibinfo {author} {\bibfnamefont {Y.}~\bibnamefont {Wang}}, \bibinfo {author} {\bibfnamefont {F.}~\bibnamefont {Guan}}, \bibinfo {author} {\bibfnamefont {X.}~\bibnamefont {Diao}}, \bibinfo {author} {\bibfnamefont {Z.}~\bibnamefont {Qin}}, \bibinfo {author} {\bibfnamefont {D.}~\bibnamefont {Si}}, \bibinfo {author} {\bibfnamefont {B.}~\bibnamefont {Zhang}}, \bibinfo {author} {\bibfnamefont {Y.}~\bibnamefont {Zhang}}, \bibinfo {author} {\bibfnamefont {B.}~\bibnamefont {Sun}}, \bibinfo {author} {\bibfnamefont {X.}~\bibnamefont {Wei}}, \bibinfo {author} {\bibfnamefont {H.}~\bibnamefont {Yang}}, \bibinfo {author} {\bibfnamefont {P.}~\bibnamefont {Ma}}, \bibinfo {author} {\bibfnamefont {H.}~\bibnamefont {Zou}}, \bibinfo {author} {\bibfnamefont {T.}~\bibnamefont {Qiu}}, \bibinfo {author} {\bibfnamefont {X.}~\bibnamefont
  {Huang}}, \bibinfo {author} {\bibfnamefont {R.}~\bibnamefont {Hu}}, \bibinfo {author} {\bibfnamefont {L.}~\bibnamefont {Duan}}, \bibinfo {author} {\bibfnamefont {F.}~\bibnamefont {Duan}}, \bibinfo {author} {\bibfnamefont {Q.}~\bibnamefont {Hu}}, \bibinfo {author} {\bibfnamefont {J.}~\bibnamefont {Ma}}, \bibinfo {author} {\bibfnamefont {S.}~\bibnamefont {Xu}}, \bibinfo {author} {\bibfnamefont {Z.}~\bibnamefont {Bai}}, \bibinfo {author} {\bibfnamefont {Y.}~\bibnamefont {Yang}},\ and\ \bibinfo {author} {\bibfnamefont {Z.}~\bibnamefont {Xiao}},\ }\bibfield  {title} {\bibinfo {title} {A csi(tl) hodoscope on cshine for bremsstrahlung γ-rays in heavy ion reactions},\ }\href {https://doi.org/https://doi.org/10.1016/j.nima.2023.168330} {\bibfield  {journal} {\bibinfo  {journal} {Nuclear Instruments and Methods in Physics Research Section A: Accelerators, Spectrometers, Detectors and Associated Equipment}\ }\textbf {\bibinfo {volume} {1053}},\ \bibinfo {pages} {168330} (\bibinfo {year} {2023})}\BibitemShut {NoStop}%
\bibitem [{\citenamefont {Guan}\ \emph {et~al.}(2021)\citenamefont {Guan}, \citenamefont {Diao}, \citenamefont {Wang}, \citenamefont {Qin}, \citenamefont {Qin}, \citenamefont {Wu}, \citenamefont {Guo}, \citenamefont {Wei}, \citenamefont {Yang}, \citenamefont {Ma}, \citenamefont {Hu}, \citenamefont {Duan}, \citenamefont {Liu}, \citenamefont {Su}, \citenamefont {Ma}, \citenamefont {Hou},\ and\ \citenamefont {Xiao}}]{GUAN2021165592}%
  \BibitemOpen
  \bibfield  {author} {\bibinfo {author} {\bibfnamefont {F.}~\bibnamefont {Guan}}, \bibinfo {author} {\bibfnamefont {X.}~\bibnamefont {Diao}}, \bibinfo {author} {\bibfnamefont {Y.}~\bibnamefont {Wang}}, \bibinfo {author} {\bibfnamefont {Y.}~\bibnamefont {Qin}}, \bibinfo {author} {\bibfnamefont {Z.}~\bibnamefont {Qin}}, \bibinfo {author} {\bibfnamefont {Q.}~\bibnamefont {Wu}}, \bibinfo {author} {\bibfnamefont {D.}~\bibnamefont {Guo}}, \bibinfo {author} {\bibfnamefont {X.}~\bibnamefont {Wei}}, \bibinfo {author} {\bibfnamefont {H.}~\bibnamefont {Yang}}, \bibinfo {author} {\bibfnamefont {P.}~\bibnamefont {Ma}}, \bibinfo {author} {\bibfnamefont {R.}~\bibnamefont {Hu}}, \bibinfo {author} {\bibfnamefont {L.}~\bibnamefont {Duan}}, \bibinfo {author} {\bibfnamefont {W.}~\bibnamefont {Liu}}, \bibinfo {author} {\bibfnamefont {W.}~\bibnamefont {Su}}, \bibinfo {author} {\bibfnamefont {C.-W.}\ \bibnamefont {Ma}}, \bibinfo {author} {\bibfnamefont {Y.}~\bibnamefont {Hou}},\ and\ \bibinfo {author} {\bibfnamefont
  {Z.}~\bibnamefont {Xiao}},\ }\bibfield  {title} {\bibinfo {title} {A compact spectrometer for heavy ion experiments in the fermi energy regime},\ }\href {https://doi.org/https://doi.org/10.1016/j.nima.2021.165592} {\bibfield  {journal} {\bibinfo  {journal} {Nuclear Instruments and Methods in Physics Research Section A: Accelerators, Spectrometers, Detectors and Associated Equipment}\ }\textbf {\bibinfo {volume} {1011}},\ \bibinfo {pages} {165592} (\bibinfo {year} {2021})}\BibitemShut {NoStop}%
\bibitem [{\citenamefont {Li}\ \emph {et~al.}(1996)\citenamefont {Li}, \citenamefont {Ren}, \citenamefont {Ko},\ and\ \citenamefont {Yennello}}]{Li:1996ix}%
  \BibitemOpen
  \bibfield  {author} {\bibinfo {author} {\bibfnamefont {B.-A.}\ \bibnamefont {Li}}, \bibinfo {author} {\bibfnamefont {Z.-Z.}\ \bibnamefont {Ren}}, \bibinfo {author} {\bibfnamefont {C.~M.}\ \bibnamefont {Ko}},\ and\ \bibinfo {author} {\bibfnamefont {S.~J.}\ \bibnamefont {Yennello}},\ }\bibfield  {title} {\bibinfo {title} {{Isospin dependence of collective flow in heavy ion collisions at intermediate-energies}},\ }\href {https://doi.org/10.1103/PhysRevLett.76.4492} {\bibfield  {journal} {\bibinfo  {journal} {Phys. Rev. Lett.}\ }\textbf {\bibinfo {volume} {76}},\ \bibinfo {pages} {4492} (\bibinfo {year} {1996})},\ \Eprint {https://arxiv.org/abs/nucl-th/9605015} {arXiv:nucl-th/9605015} \BibitemShut {NoStop}%
\bibitem [{\citenamefont {Li}\ \emph {et~al.}(1997)\citenamefont {Li}, \citenamefont {Ko},\ and\ \citenamefont {Ren}}]{Li:1997rc}%
  \BibitemOpen
  \bibfield  {author} {\bibinfo {author} {\bibfnamefont {B.-A.}\ \bibnamefont {Li}}, \bibinfo {author} {\bibfnamefont {C.~M.}\ \bibnamefont {Ko}},\ and\ \bibinfo {author} {\bibfnamefont {Z.-z.}\ \bibnamefont {Ren}},\ }\bibfield  {title} {\bibinfo {title} {{Equation of state of asymmetric nuclear matter and collisions of neutron rich nuclei}},\ }\href {https://doi.org/10.1103/PhysRevLett.78.1644} {\bibfield  {journal} {\bibinfo  {journal} {Phys. Rev. Lett.}\ }\textbf {\bibinfo {volume} {78}},\ \bibinfo {pages} {1644} (\bibinfo {year} {1997})},\ \Eprint {https://arxiv.org/abs/nucl-th/9701048} {arXiv:nucl-th/9701048} \BibitemShut {NoStop}%
\bibitem [{\citenamefont {Li}\ \emph {et~al.}(2004)\citenamefont {Li}, \citenamefont {Das}, \citenamefont {Das~Gupta},\ and\ \citenamefont {Gale}}]{Li:2003ts}%
  \BibitemOpen
  \bibfield  {author} {\bibinfo {author} {\bibfnamefont {B.-A.}\ \bibnamefont {Li}}, \bibinfo {author} {\bibfnamefont {C.~B.}\ \bibnamefont {Das}}, \bibinfo {author} {\bibfnamefont {S.}~\bibnamefont {Das~Gupta}},\ and\ \bibinfo {author} {\bibfnamefont {C.}~\bibnamefont {Gale}},\ }\bibfield  {title} {\bibinfo {title} {{Effects of momentum dependent symmetry potential on heavy ion collisions induced by neutron rich nuclei}},\ }\href {https://doi.org/10.1016/j.nuclphysa.2004.02.016} {\bibfield  {journal} {\bibinfo  {journal} {Nucl. Phys. A}\ }\textbf {\bibinfo {volume} {735}},\ \bibinfo {pages} {563} (\bibinfo {year} {2004})},\ \Eprint {https://arxiv.org/abs/nucl-th/0312054} {arXiv:nucl-th/0312054} \BibitemShut {NoStop}%
\bibitem [{\citenamefont {Li}\ \emph {et~al.}(2018)\citenamefont {Li}, \citenamefont {Cai}, \citenamefont {Chen},\ and\ \citenamefont {Xu}}]{LI201829}%
  \BibitemOpen
  \bibfield  {author} {\bibinfo {author} {\bibfnamefont {B.-A.}\ \bibnamefont {Li}}, \bibinfo {author} {\bibfnamefont {B.-J.}\ \bibnamefont {Cai}}, \bibinfo {author} {\bibfnamefont {L.-W.}\ \bibnamefont {Chen}},\ and\ \bibinfo {author} {\bibfnamefont {J.}~\bibnamefont {Xu}},\ }\bibfield  {title} {\bibinfo {title} {Nucleon effective masses in neutron-rich matter},\ }\href {https://doi.org/https://doi.org/10.1016/j.ppnp.2018.01.001} {\bibfield  {journal} {\bibinfo  {journal} {Progress in Particle and Nuclear Physics}\ }\textbf {\bibinfo {volume} {99}},\ \bibinfo {pages} {29} (\bibinfo {year} {2018})}\BibitemShut {NoStop}%
\bibitem [{\citenamefont {Hen}\ \emph {et~al.}(2015)\citenamefont {Hen}, \citenamefont {Li}, \citenamefont {Guo}, \citenamefont {Weinstein},\ and\ \citenamefont {Piasetzky}}]{Hen:2014yfa}%
  \BibitemOpen
  \bibfield  {author} {\bibinfo {author} {\bibfnamefont {O.}~\bibnamefont {Hen}}, \bibinfo {author} {\bibfnamefont {B.-A.}\ \bibnamefont {Li}}, \bibinfo {author} {\bibfnamefont {W.-J.}\ \bibnamefont {Guo}}, \bibinfo {author} {\bibfnamefont {L.~B.}\ \bibnamefont {Weinstein}},\ and\ \bibinfo {author} {\bibfnamefont {E.}~\bibnamefont {Piasetzky}},\ }\bibfield  {title} {\bibinfo {title} {{Symmetry Energy of Nucleonic Matter With Tensor Correlations}},\ }\href {https://doi.org/10.1103/PhysRevC.91.025803} {\bibfield  {journal} {\bibinfo  {journal} {Phys. Rev. C}\ }\textbf {\bibinfo {volume} {91}},\ \bibinfo {pages} {025803} (\bibinfo {year} {2015})},\ \Eprint {https://arxiv.org/abs/1408.0772} {arXiv:1408.0772 [nucl-ex]} \BibitemShut {NoStop}%
\bibitem [{\citenamefont {Gan}\ \emph {et~al.}(1994)\citenamefont {Gan}, \citenamefont {Brinkmann}, \citenamefont {Caraley}, \citenamefont {Fineman}, \citenamefont {Kernan}, \citenamefont {McGrath},\ and\ \citenamefont {Danielewicz}}]{Gan1994298}%
  \BibitemOpen
  \bibfield  {author} {\bibinfo {author} {\bibfnamefont {N.}~\bibnamefont {Gan}}, \bibinfo {author} {\bibfnamefont {K.-T.}\ \bibnamefont {Brinkmann}}, \bibinfo {author} {\bibfnamefont {A.}~\bibnamefont {Caraley}}, \bibinfo {author} {\bibfnamefont {B.}~\bibnamefont {Fineman}}, \bibinfo {author} {\bibfnamefont {W.}~\bibnamefont {Kernan}}, \bibinfo {author} {\bibfnamefont {R.}~\bibnamefont {McGrath}},\ and\ \bibinfo {author} {\bibfnamefont {P.}~\bibnamefont {Danielewicz}},\ }\bibfield  {title} {\bibinfo {title} {Neutron-proton bremsstrahlung from low-energy heavy-ion reactions},\ }\href {https://doi.org/10.1103/PhysRevC.49.298} {\bibfield  {journal} {\bibinfo  {journal} {Physical Review C}\ }\textbf {\bibinfo {volume} {49}},\ \bibinfo {pages} {298 } (\bibinfo {year} {1994})}\BibitemShut {NoStop}%
\bibitem [{\citenamefont {Si}\ \emph {et~al.}(2024)\citenamefont {Si} \emph {et~al.}}]{Si:2023dmz}%
  \BibitemOpen
  \bibfield  {author} {\bibinfo {author} {\bibfnamefont {D.-W.}\ \bibnamefont {Si}} \emph {et~al.},\ }\bibfield  {title} {\bibinfo {title} {{Measurement of the high energy \ensuremath{\gamma}-rays from heavy ion reactions using \v{C}erenkov detector}},\ }\href {https://doi.org/10.1007/s41365-024-01368-7} {\bibfield  {journal} {\bibinfo  {journal} {Nucl. Sci. Tech.}\ }\textbf {\bibinfo {volume} {35}},\ \bibinfo {pages} {24} (\bibinfo {year} {2024})},\ \Eprint {https://arxiv.org/abs/2307.12995} {arXiv:2307.12995 [physics.ins-det]} \BibitemShut {NoStop}%
\end{thebibliography}%


\appendix

\newpage
\section{Original $\gamma$ Spectrum}\label{OrigSpecTable}

\begin{table}[htbp]
\caption{\label{Dataoutput} The original $\gamma$ spectrum (in CM reference frame) reconstructed by RL algorithm.}
\begin{ruledtabular}
\begin{tabular}{ccc}
$E_{\gamma} (\rm MeV)$ &$\frac{d^2P_{\gamma}}{dE_{\gamma}d\Omega} (\rm MeV^{-1}\cdot \rm sr^{-1})$ &$\sigma (\rm MeV^{-1}\cdot \rm sr^{-1})$ \\
\hline
3.6 &$3.04\times10^{-1}$ &$2.05\times10^{-4}$\\
8.8 &$3.43\times10^{-2}$ &$1.40\times10^{-4}$ \\
14.0 &$1.78\times10^{-3}$ &$3.73\times10^{-5}$ \\
19.2 &$2.74\times10^{-4}$ &$1.43\times10^{-5}$ \\
24.4 &$7.32\times10^{-5}$ &$7.73\times10^{-6}$ \\
29.5 &$4.07\times10^{-5}$ &$5.02\times10^{-6}$ \\
34.7 &$2.83\times10^{-5}$ &$4.34\times10^{-6}$ \\
39.9 &$6.97\times10^{-6}$  &$3.14\times10^{-6}$ \\
45.1 &$1.61\times10^{-5}$ &$3.61\times10^{-6}$ \\
50.3 &$6.79\times10^{-6}$ &$3.02\times10^{-6}$ \\
55.5 &$1.53\times10^{-6}$ &$2.08\times10^{-6}$ \\
60.6 &$6.42\times10^{-6}$ &$2.41\times10^{-6}$ \\
65.8 &$4.82\times10^{-6}$ &$2.72\times10^{-6}$ \\
71.0 &$3.10\times10^{-6}$ &$2.53\times10^{-6}$ \\
76.2 &$2.38\times10^{-7}$ &$1.07\times10^{-6}$ \\
81.4 &$4.20\times10^{-6}$ &$2.85\times 10^{-6}$ \\
\end{tabular}
\end{ruledtabular}
\end{table}

\newpage

\section{Experimental $\gamma$ spectrum}\label{ExpSpecTable}

\begin{table}[htbp]
\centering
\caption{\label{experimentData} Experiment measurement spectrum (in laboratory reference frame).}
\begin{ruledtabular}
\begin{tabular}{cc|cc|cc|cc}
$E_{\gamma}$\footnote{The unit of $E_{\gamma}$ is $\rm MeV$.} &$\frac{dN_{\gamma}}{dE_{\gamma}}$\footnote{The unit of $\frac{dN_{\gamma}}{dE_{\gamma}}$ is $\rm MeV^{-1}$.} &$E_{\gamma}$ &$\frac{dN_{\gamma}}{dE_{\gamma}}$ &$E_{\gamma}$ &$\frac{dN_{\gamma}}{dE_{\gamma}}$ &$E_{\gamma}$ &$\frac{dN_{\gamma}}{dE_{\gamma}}$\\
\hline
1 & 467657 &21 & 59 &41 & 7 &61 & 2  \\
2 & 179002 &22 & 43 &42 & 5 &62 & 6  \\
3 & 98944 &23 & 53 &43 & 6 &63 & 1  \\
4 & 63378 &24 & 27 &44 & 0 &64 & 3  \\
5 & 47859 &25 & 31 &45 & 4 &65 & 2  \\
6 & 39592 &26 & 35 &46 & 2 &66 & 0  \\
7 & 24209 &27 & 28 &47 & 4 &67 & 2  \\
8 & 10594 &28 & 33 &48 & 4 &68 & 1  \\
9 & 4888 &29 & 13 &49 & 1 &69 & 0  \\
10 & 2596 &30 & 23 &50 & 3 &70 & 1  \\
11 & 1425 &31 & 20 &51 & 2 &71 & 0  \\
12 & 904 &32 & 11 &52 & 2 &72 & 0  \\
13 & 621 &33 & 18 &53 & 3 &73 & 0  \\
14 & 385 &34 & 12 &54 & 2 &74 & 0  \\
15 & 281 &35 & 8 &55 & 6 &75 & 0  \\
16 & 195 &36 & 12 &56 & 5 &76 & 0  \\
17 & 145 &37 & 8 &57 & 0 &77 & 1  \\
18 & 99 &38 & 6 &58 & 2 &78 & 0  \\
19 & 94 &39 & 14 &59 & 2 &79 & 0  \\
20 & 64 &40 & 12 &60 & 3 &80 & 0  \\
\end{tabular}
\end{ruledtabular}
\end{table}


\end{document}